\newif\ifonecol

\onecoltrue    

\ifonecol
\documentclass[11pt,draftcls,onecolumn]{IEEEtran}
\else
\documentclass[journal]{IEEEtran}
\fi
\IEEEoverridecommandlockouts
\usepackage{amsmath,epsfig,amssymb,amsthm,url}
\usepackage{algorithm}
\usepackage{algpseudocode}
\usepackage{multirow}
\usepackage{mdframed}
\usepackage{url}
\usepackage{array}
\usepackage[multiple]{footmisc}

\newcommand{\tr}{ {\mbox{trace}} }

\newcommand{\Rbb}{{\mathbb{R}}}
\newcommand{\Cbb}{{\mathbb{C}}}

\newcommand{\Ab}{{\bf A}}

\newcommand{\Cb}{{\bf C}}

\newcommand{\Pb}{{\bf P}}

\newcommand{\xb}{{\bf x}}

\newcommand{\yb}{{\bf y}}

\newcommand{\Zb}{{\bf Z}}

\newcommand{\Db}{{\bf D}}
\newcommand{\Bb}{{\bf B}}

\newcommand{\Xb}{{\bf X}}
\newcommand{\Yb}{{\bf Y}}

\newcommand{\soft}[2]{{\cal S}_{#2}({#1})}
\newcommand{\nz}[1]{\|#1\|_0} 
\newcommand{\no}[1]{\|#1\|_1} 
\newcommand{\nf}[1]{\|#1\|_F} 

\newcommand{\Psib}{{\mbox{\boldmath $\Psi$}}}

\newcommand{\sign}{\mbox{{sign}}}

\newcommand{\sbt}{\mbox{{subject~to}}}

\newcommand{\Ib}{{\bf I}}

\newcommand{\Lamb}{{\mbox{\boldmath $\Lambda$}}}

\newcommand{\Ub}{{\mathbf U}}
\newcommand{\Fb}{{\mathbf F}}

\newcommand{\Vb}{{\mathbf V}}

\newsavebox\mybox

\begin{document}
\title{A Fast and Efficient Algorithm for Reconstructing MR images From Partial Fourier Samples}

\author{Fateme~Ghayem$^{*}$,~Farokh~Marvasti,~\IEEEmembership{Senior Member,~IEEE}
\thanks{The authors are with the Electrical Engineering Department, Sharif University of Technology, Tehran, Iran. fateme.ghayem@gmail.com,
fmarvasti@gmail.com}}

\maketitle

\begin{abstract}

In this paper, the problem of Magnetic Resonance (MR) image reconstruction from partial Fourier samples has been considered. To this aim, we leverage the evidence that MR images are sparser than their zero-filled reconstructed ones from incomplete Fourier samples. This information can be used to define an optimization problem which searches for the sparsest possible image conforming with the available Fourier samples. We solve the resulting problem using the well-known Alternating Direction Method of Multipliers (ADMM). Unlike most existing methods that work with small over-lapping image patches, the proposed algorithm considers the whole image without dividing it into small blocks. Experimental results prominently confirm its promising performance and advantages over the existing methods.

\end{abstract}


\IEEEpeerreviewmaketitle

\section{INTRODUCTION}
The pervasive impact of the Magnetic Resonance Imaging (MRI) as a competent approach to investigate the anatomy and physiology of the body can not be ignored in the present context of medical imaging. In MRI, the data are acquired as partial Fourier samples in k-space. The goal is to accurately reconstruct the medical images from these incomplete samples jointly assuring for the minimal scan time. 

Many techniques have already been proposed to reconstruct images using as fewer samples as possible, and hence reducing the scan time \cite{LianL99, AggaB08}. One such state of the art approach is to use the theory of compressed sensing (CS) which demonstrates highly promising results \cite{LustDSP08, Char09}. For this, the overall reconstruction is formulated as a sparse coding problem, in which the original image is assumed to have a sparse representation in an appropriate domain like wavelets, Discrete Cosine Transform (DCT), contourlets, etc., while conforming with the available compressed measurements (i.e., partial Fourier samples). The basic problem formulation is as follows
\begin{equation}
\min_{\xb}~\nz{\Psib\xb}~~~\sbt~~~\Fb_u\xb=\yb,
\end{equation}
where $\xb$ is the vectorized version of the estimate of the original image, $\nz{.}$ denotes the $\ell_0$ pseudo norm which counts the number of nonzero elements, $\Psib$ is some sparsifying basis like DCT, $\Fb_u$ is an undersampled Fourier matrix, and $\yb$ is the corresponding available measurements (Fourier samples). Numerous algorithms exist which aim to solve this problem or its noise-aware variant and get approximate solutions, see e.g., Iterative Method with Adaptive Thresholding (IMAT) \cite{AzghM13, MarvA12, BoloAEKM15, IMAT15}, Smoothed L0 norm (SL0) \cite{MohiBJ09}, Orthogonal Matching Pursuit (OMP) \cite{PatiRK93}.

In \cite{RaviB11}, the authors proposed to use \textit{adaptive} sparsifying transforms instead of fixed ones like DCT and wavelets. The idea of \textit{dictionary learning} in sparse signal processing \cite{SadeBJ13, SadeBJ14, RubiBE10} is adopted to simultaneously learn the sparsifying transform and reconstruct the image. The experimental results show significant improvements over previous algorithms.
\begin{figure}[t]
\label{fig:sp}
\centering
\centerline{\includegraphics[scale=0.25]{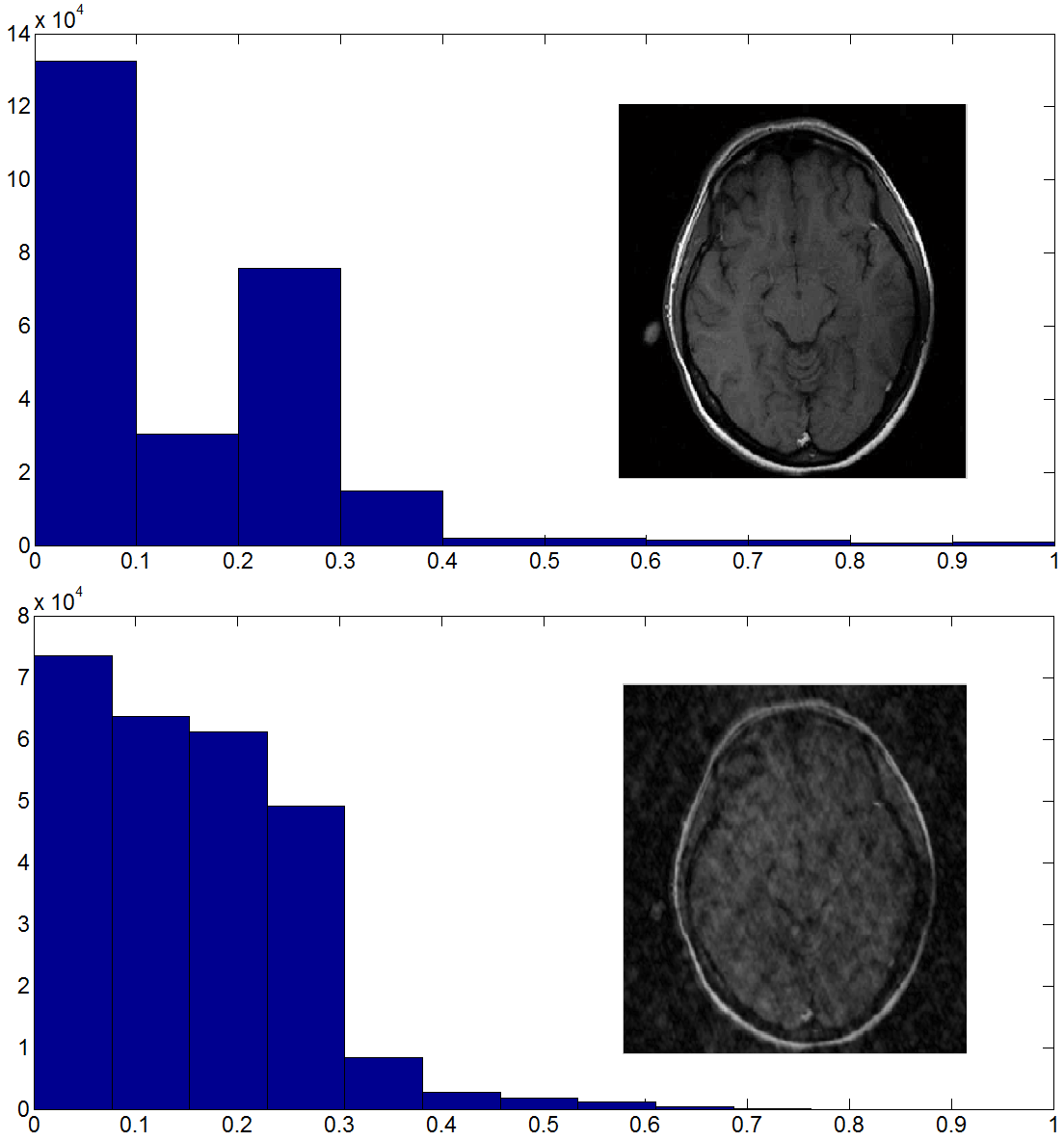}}
\caption{Histograms of original (top) and zero-filled reconstructed (bottom).}
\end{figure}

In this paper, we propose a simple yet fast and efficient algorithm for the reconstruction of MRI images. The idea is based on the evidence that MR images are sparser than their zero-filled reconstructed ones, i.e., the ones obtained as the inverse Fourier transform of the available Fourier samples by replacing the unknown samples with zeroes. This is illustrated in Fig.~1. In this figure, the histograms of one typical MR image and that of its zero-filled reconstruction are shown. As can be seen, the original image is sparser than its zero-filled reconstruction. This motivates us to formulate the image reconstruction problem as looking for the sparsest possible image conforming with the available Fourier samples. We solve it using the well-known Alternating Direction Method of Multipliers (ADMM) method. We work with the whole image instead of small overlapping blocks as in \cite{RaviB11}. Our experimental simulations indicate improvements in both running time and quality of the reconstructed images, as compared to \cite{RaviB11}.

The rest of the paper is organized as follows. In Section~\ref{sec:prop} we formulate the problem. This is followed by a review on ADMM and applying it on our own problem. Simulation results are presented in Section~\ref{sec:sim}. Finally, concluding remarks are presented in Section~\ref{sec:conc}.
\section{PROBLEM FORMULATION}\label{sec:prop}
Consider an image $\Xb_0\in\Rbb^{N\times N}$. There are only a subset of its Fourier samples indexed in $\Omega$ where $|\Omega|=M$ with $M$ usually being much smaller than the whole samples, i.e., $N\times N=N^2$. The objective is to reconstruct the original image $\Xb_0$ from its incomplete Fourier samples in $\Omega$. Hence, we exploit the evidence which was discussed in previous section and illustrated in Fig.~1. To measure the sparsity, we use the $\ell_1$ norm $\no{\Xb}$, which is defined as the sum of the absolute values of the entries in $\Xb$.

Consider the Discrete Fourier Transform (DFT) matrix $\Db\in\Cbb^{N\times N}$. From the separability of DFT, the 2D DFT of an image $\Xb$ is defined as $\Yb=\Db\Xb\Db^H$, where ``$H$" denotes conjugate transpose\footnote{Note that there is a normalization factor of $M$ in $\Db^H$ which we have omitted for simplicity.}. The inverse DFT is simply $\Xb=\Db^H\Yb\Db$. Using this, we propose the following problem to reconstruct the original image $\Xb_0$ ($\Xb_0=\mbox{IFFT}(\Yb)$)
\begin{equation}
\label{eq:propp}
\min_{\Yb}~\no{\Db^H\Yb\Db}~\sbt~\Yb(\Omega)=\Yb_0(\Omega).
\end{equation}
Note that the noise-aware variant of the above problem is as follows, which we do not discuss in this paper,
\begin{equation}
\label{eq:proppna}
\min_{\Yb}~\no{\Db^H\Yb\Db}+\lambda\nf{\Yb(\Omega)-\Yb_0(\Omega)}^2.
\end{equation}
To solve \eqref{eq:propp}, we use the well-known Alternating Direction Method of Multipliers (ADMM). In priority, we first review the main points of ADMM.
\subsection{ADMM}
Consider the following equality-constrained
problem
\begin{equation}
\label{eq:admm}
\min_{\Ub, \Vb} f_1(\Ub)+f_2(\Vb) ~\sbt~ \Ab\Ub+\Bb\Vb=\Cb,
\end{equation}
where $\Ub$ and $\Vb$ are optimization variables (not necessarily of the same size), $f_1$  and $f_2$ are convex functions, and $\Ab$, $\Bb$, and $\Cb$ are the problem parameters. A common approach to solve such problems is to use the notion of augmented Lagrangian function, which is defined as follows
\begin{equation}
L_{\mu}(\Ub,\Vb,\Lamb)=f_1(\Ub)+f_2(\Vb)-\tr(\Lamb^T(\Ab\Ub+\Bb\Vb-\Cb))+\frac{\mu}{2}\nf{\Ab\Ub+\Bb\Vb-\Cb}^2,
\end{equation}
where, $\Lamb$ is the Lagrangian multipliers matrix associated to the equality constraint and $\mu$ is a positive parameter. The idea of an ADMM solver is to iteratively minimize the above augmented Lagrangian function over $\Ub$ and $\Vb$, along with update of $\Lamb$. 
\subsection{Algorithm}
To solve \eqref{eq:propp} using ADMM, we reformulate it as follows
\begin{equation}
\min_{\Zb,\Yb}~\no{\Zb}~\sbt~\Yb(\Omega)=\Yb_0(\Omega),~\Zb=\Db^H\Yb\Db,
\end{equation}
where $\Zb$ is an axillary variable of the same size as $\Yb$. The Lagrangian function for this problem is
\begin{multline}
\label{eq:lf}
L_{\mu}(\Zb,\Yb,\Lamb_1,\Lamb_2)=\no{\Zb}-\tr(\Lamb_1^T(\Yb(\Omega)-\Yb_0(\Omega))-\tr(\Lamb_2^T(\Zb-\Db^H\Yb\Db))\\
+\frac{\mu_1}{2}\nf{\Yb(\Omega)-\Yb_0(\Omega))}^2+\frac{\mu_2}{2}\nf{\Zb-\Db^H\Yb\Db}^2,
\end{multline}
The $\Zb$-update problem, after some straightforward calculations becomes
\begin{equation}
\min_{\Zb}~\no{\Zb}+\frac{\mu_2}{2}\nf{\Zb-\Db^H\Yb\Db-\frac{1}{\mu_2}\Lamb_2}^2,
\end{equation}
This problem is separable over entries of $\Zb$, which ends up with $N^2$ scalar sub-problems of the form
\begin{equation}
\label{eq:l1}
\min_{x}~|x|+\frac{\mu_2}{2}(x-a)^2
\end{equation}
where $x$ and $a$ denote typical entries of $\Zb$ and $\Db^H\Yb\Db+\frac{1}{\mu_2}\Lamb_2$, respectively. Problem~\eqref{eq:l1} is well-known in the compressed sensing community, which has the following simple closed-form solution
\begin{equation}
x^*=\soft{a}{\frac{1}{\mu_2}},
\end{equation}
where $\soft{a}{\lambda}$ is the soft-thresholding function defined as
\begin{equation}
\soft{a}{\lambda}\triangleq \sign(a)\cdot\max(|a|-\lambda,0).
\end{equation}
Final update formula for $\Zb$ is
\begin{equation}
\label{eq:zup}
\Zb=\soft{\Db^H\Yb\Db+\frac{1}{\mu_2}\Lamb_2}{\frac{1}{\mu_2}},
\end{equation}
in which, the soft-thresholding function acts component wise.

To update $\Yb$, after re-arranging the terms in \eqref{eq:lf}, we have the following problem
\begin{equation}
\label{eq:xup1}
\min_{\Yb}~\frac{\mu_1}{2}\nf{\Yb(\Omega)-\Yb_0(\Omega)-\frac{1}{\mu_1}\Lamb_1}^2+\frac{\mu_2}{2}\nf{\Zb-\Db^H\Yb\Db-\frac{1}{\mu_2}\Lamb_2}^2.
\end{equation}
We use the orthonormal property of the DFT matrix, i.e., $\Db^H\Db=\Ib$, to rewrite problem~\eqref{eq:xup1} into the following equivalent form
\begin{equation}
\label{eq:xup2}
\min_{\Yb}~\frac{\mu_1}{2}\nf{\Yb(\Omega)-\Yb_0(\Omega)-\frac{1}{\mu_1}\Lamb_1}^2+\frac{\mu_2}{2}\nf{\Yb-\Db(\Zb-\frac{1}{\mu_2}\Lamb_2)\Db^H}^2.
\end{equation}
Setting the gradient of the above objective function with respect to $\Yb$ equal to $0$ results in
\begin{equation}
\label{eq:xup3}
\mu_1\Pb.*(\Yb(\Omega)-\Yb_0(\Omega)-\frac{1}{\mu_1}\Lamb_1)+\mu_2(\Xb-\Db(\Zb-\frac{1}{\mu_2}\Lamb_2)\Db^H)=0,
\end{equation}
where $\Pb$ is a binary matrix having $1$ in the entries in $\Omega$ and $0$ elsewhere and $.*$ denotes component wise multiplication.

Let us define 
\begin{equation}
\label{eq:adef}
\Ab\triangleq \Db(\Zb-\frac{1}{\mu_2}\Lamb_2)\Db^H,
\end{equation}
and
\begin{equation}
\label{eq:bup}
\Bb\triangleq \frac{\mu_1(\Yb_0+\frac{1}{\mu_1}\Lamb_1)+\mu_2\Ab}{\mu_1+\mu_2}.
\end{equation}
From \eqref{eq:xup3}, we have
\begin{equation}
\label{eq:xupf}
\Yb(i,j)=\left\{
\begin{array}{rl}
\Bb(i,j) & ~~~\in \Omega\\
\Ab(i,j) &~~~\not\in \Omega
\end{array} .\right.
\end{equation}
The formula for updating the Lagrangian multipliers matrices are as follows
\begin{equation}
\label{eq:l1up}
\Lamb_1=\Lamb_1-\mu_1(\Yb(\Omega)-\Yb_0(\Omega)),
\end{equation}
\begin{equation}
\label{eq:l2up}
\Lamb_2=\Lamb_2-\mu_2(\Zb-\Db^H\Yb\Db),
\end{equation}
\subsection{Fast implementation}
Note that the matrix multiplications $\Db\Zb\Db^H$ and $\Db^H\Yb\Db$ can be efficiently implemented using the Fast Fourier Transform (FFT) of $\Zb$ and the inverse FFT (IFFT) of $\Yb$, respectively. For example, instead of \eqref{eq:adef}, we write
\begin{equation}
\label{eq:upafft}
\Ab\triangleq \mbox{FFT}(\Zb-\frac{1}{\mu_2}\Lamb_2),
\end{equation}
and instead of \eqref{eq:l2up} we write
\begin{equation}
\Lamb_2=\Lamb_2-\mu_2(\Zb-\mbox{IFFT}(\Yb)),
\end{equation}
The main steps of the final proposed algorithm are summarized in Alg.~\ref{alg:prop}.

\begin{algorithm}[t]
\caption{Proposed algorithm.}
\label{alg:prop}
\begin{algorithmic}[]
\State \textbf{Require:} $\Yb_0(\Omega)$, $\mu_1$ and $\mu_2$.
\State \textbf{Initialization:} $\Lamb_1=\Lamb_2=0$
\For{$iter=1, 2, \hdots$}
\State $\Zb=\soft{\mbox{IFFT}(\Yb)+\frac{1}{\mu_2}\Lamb_2}{\frac{1}{\mu_2}}$
\State Update $\Yb$ using \eqref{eq:xupf} along with \eqref{eq:upafft} and \eqref{eq:bup}.
\State $\Lamb_1=\Lamb_1-\mu_1(\Yb(\Omega)-\Yb_0(\Omega))$
\State $\Lamb_2=\Lamb_2-\mu_2(\Zb-\mbox{IFFT}(\Yb))$
\EndFor
\State \textbf{Output:} $\Xb=\mbox{IFFT}(\Yb)$
\end{algorithmic}
\end{algorithm}

\section{EXPERIMENTS}\label{sec:sim}
\begin{figure*}[t]
\label{fig:sp}
\begin{center}
{\includegraphics[scale=.6]{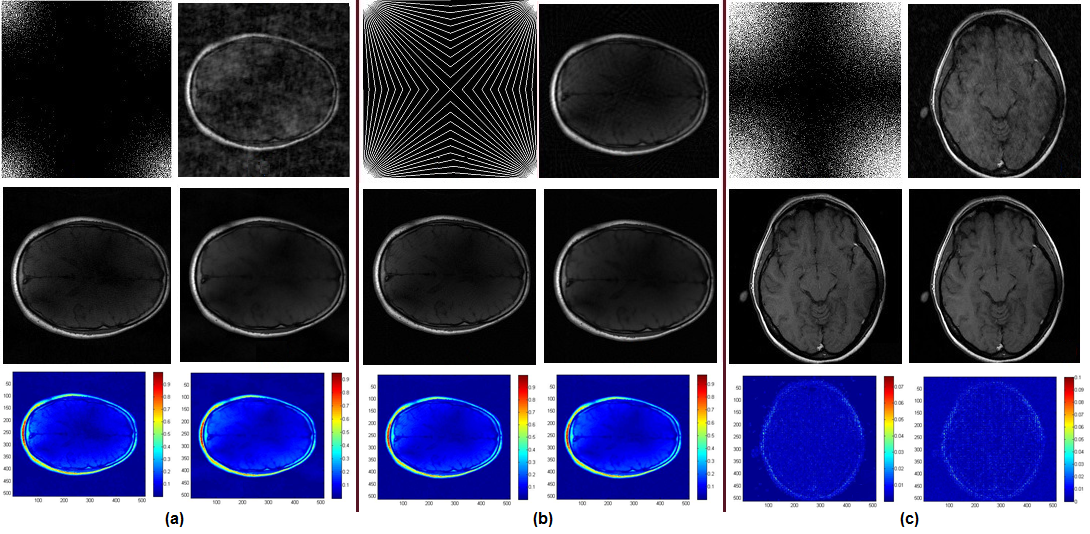}}
\end{center}
\caption{Sample results of the algorithms. There are three separate sections, i.e., (a)-(c). In each section, from left to right and top to bottom the images correspond to: the used mask, the zero-filled image, output of the proposed algorithm, output of the DLMRI algorithm, reconstruction error magnitude for our algorithm, and reconstruction error magnitude for DLMRI. (a) Random (5\%), the vivo image, (b) Radial (16\%), the vivo image, and (c) Random (25\%), the vivo image.}
\end{figure*}
In this section, we evaluate the performance of the proposed algorithm. Since a comparison to all other methods is not possible, so we chose the method in \cite{RaviB11} as an example. MATLAB R2014a on a system with an Intel Core i5, 2.27 GHz CPU, 4 GB memory and 68-bit windows 8 was employed to run the codes. The MATLAB codes, reference image and masks used in \cite{RaviB11} are available at the second author's web page\footnote{\url{http://www.ifp.illinois.edu/~yoram/DLMRI-Lab/Documentation.html}}. 

For comparison, both the algorithms (i.e., the DLMRI and our proposed one) were subjected to two different datasets, i.e., the test MR image used in \cite{RaviB11} and another MR image, which we name as MRI-1 and MRI-2, respectively. These data are depicted in Fig.~2. Various sampling masks i.e., Random, Cartesian and pseudo radial available in \cite{RaviB11} were applied on the reference data in k-space domain to reach different levels of undersamplings. It is noticeable that in pseudo radial mask, a $512\times 512$ Cartesian grid is used to map the samples on the radial lines to their nearest points on the grid.

We considered the time it takes to attain convergence in both the algorithms. To evaluate reconstruction qualities, as in \cite{RaviB11} we used Peak Signal to Noise Ratio (PSNR) in dB which is computed as the ratio of the peak intensity value of the reference image to the root mean square (RMS) reconstruction error relative to the reference image. 

The parameters of DLMRI were set according to \cite{RaviB11}. For the proposed algorithm, we fixed the parameter $\mu_1=10$, as it hardly affects the results significantly. The parameter $\mu_2$ is also scaled in the range of 10 to 30, so as to opt the best value for each mask. Subsequently, one of the advantages of our algorithm over DLMRI is its fewer number of parameters.

Table~I shows the results on both test images. For comparison, the run time, start up PSNR, i.e., PSNR(init.), and its value after reconstruction, i.e., PSNR(end) have been reported. It can be seen that our algorithm is significantly faster than DLMRI. Moreover, it outperforms DLMRI by 2-5 dB, mostly in case of random sampling.
\begin{table}[t]
\label{tbl:res}
\begin{center}
\caption{PSNR in dB and running times. In each cell, top: our algorithm and bottom: DLMRI \cite{RaviB11}.}
\centerline{\includegraphics[scale=0.8]{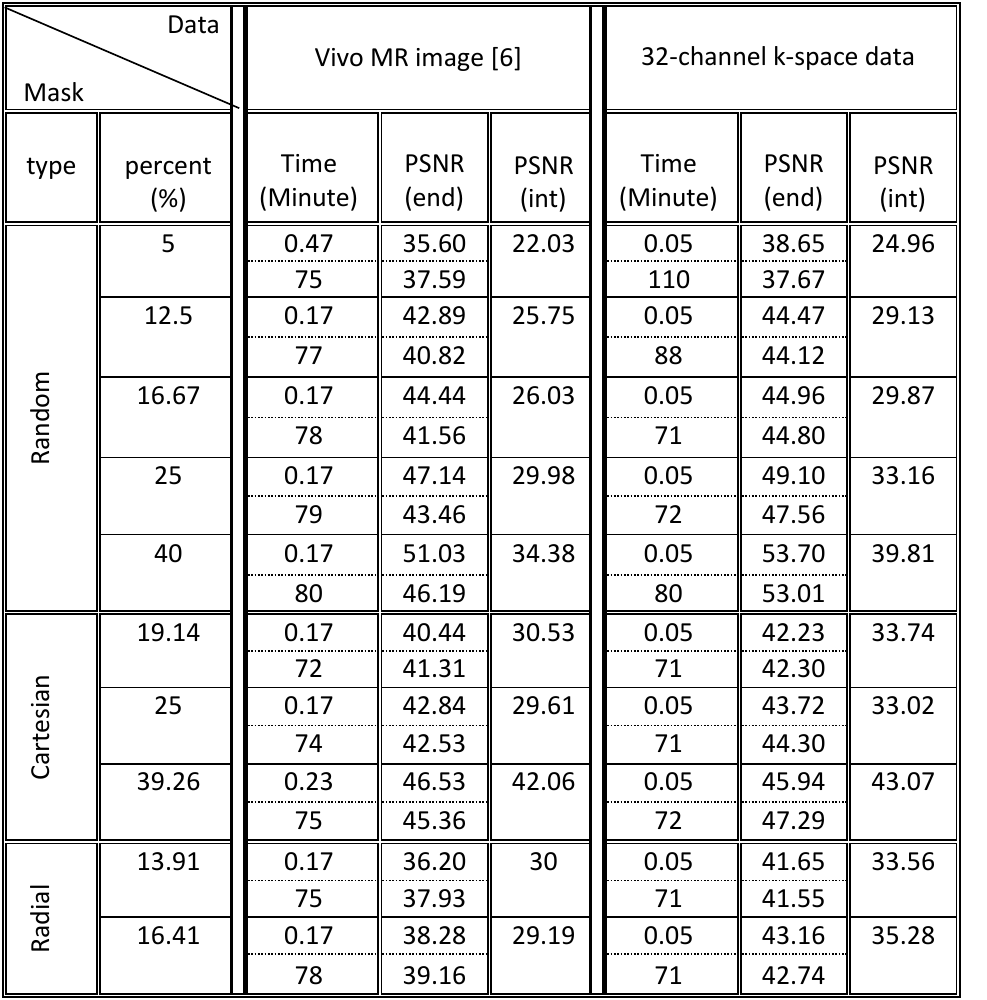}}
\end{center}
\end{table}

Figure 2 shows the sampler mask, zero-filled image, reconstructed image using DLMRI and our algorithm, and the reconstruction error magnitude images for some sample situations. It can be seen that DLMRI smooths out the images while our proposed algorithm preserves the details. The PSNR versus iterations are also demonstrated in Fig.~4. This figure shows that the convergence rate of our algorithm is higher than that of DLMRI.
\section{CONCLUSION AND FUTURE WORK}\label{sec:conc}
In this paper, we proposed a new reconstruction algorithm for MR images from their partial Fourier samples, which is governed by the sparsity of the original image. We formulate it as a minimization problem and solved it using ADMM.  The resulting algorithm has a fast implementation using FFT. Also, its another important advantage over state of the art methods like DLMRI \cite{RaviB11} is its fewer free parameters. Experimental results further confirm our arguments.

For future work, extension of the algorithm to noisy case, improving the solutions using other image priors and regularizations like total variation, and applying adaptive sparsifying transforms are of interest.
\begin{figure}[t]
\label{fig:img}
\centering
\centerline{\includegraphics[scale=0.6]{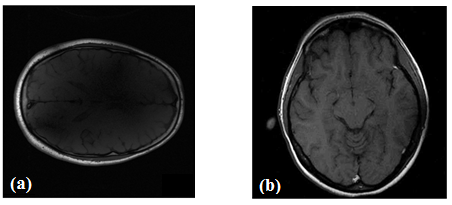}}
\caption{Test data. (a) the 32-channel k-space, (b) the vivo.}
\end{figure}
\begin{figure}[h!]
\label{fig:psnrs}
\centering
\centerline{\includegraphics[scale=0.3]{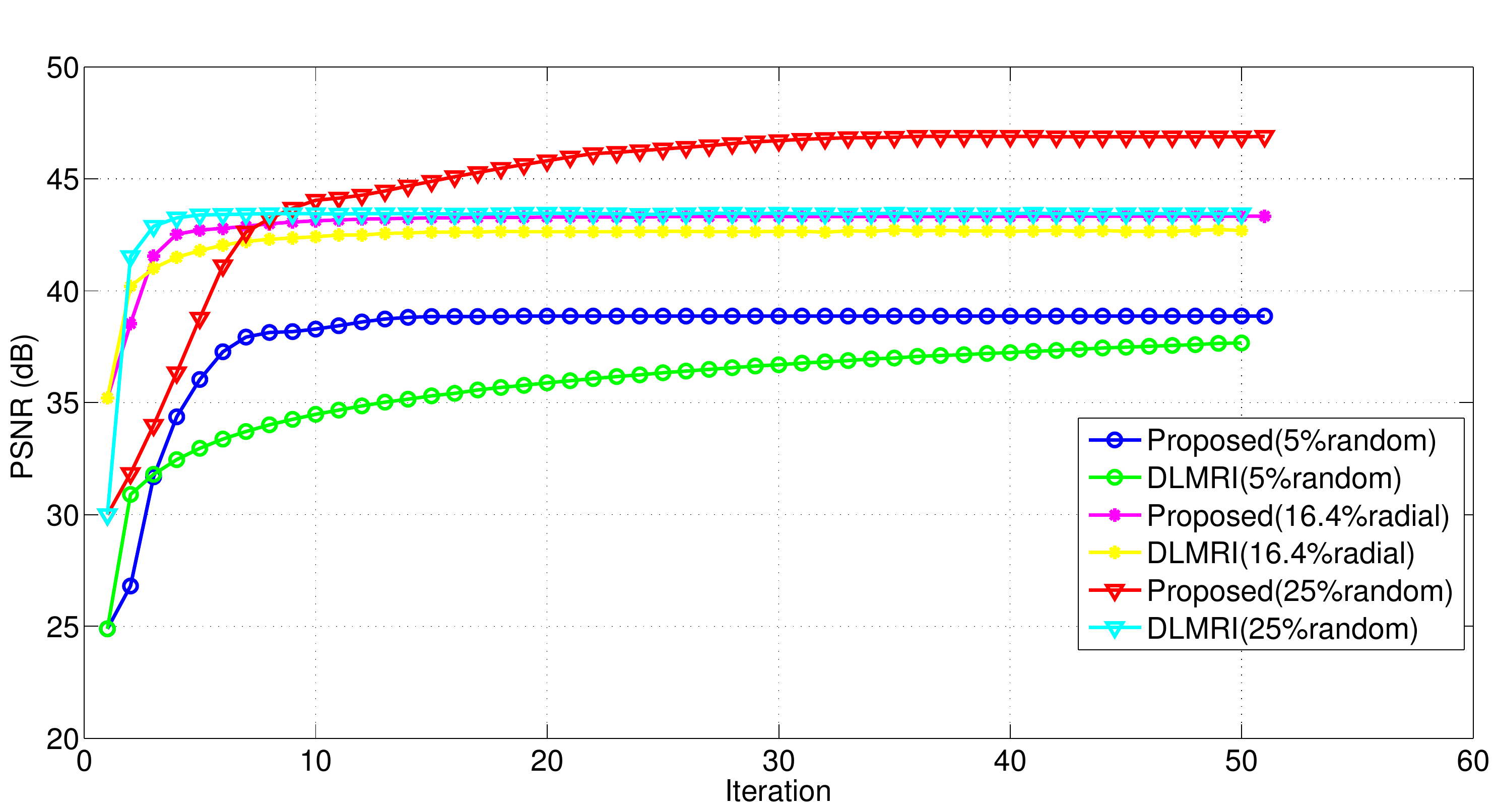}}
\caption{PSNR (dB) versus iteration number.}
\end{figure}
\bibliographystyle{IEEEbib}
\bibliography{msgnref}
\end{document}